# P3DFFT: a framework for parallel computations of Fourier transforms in three dimensions

Dmitry Pekurovsky

San Diego Supercomputer Center

UC San Diego

**Abstract.** Fourier and related transforms is a family of algorithms widely employed in diverse areas of computational science, notoriously difficult to scale on high-performance parallel computers with large number of processing elements (cores). This paper introduces a popular software package called P3DFFT implementing Fast Fourier Transforms (FFT) in three dimensions (3D) in a highly efficient and scalable way. It overcomes a well-known scalability bottleneck of 3D FFT implementations by using two-dimensional domain decomposition. Designed for portable performance, P3DFFT achieves excellent timings for a number of systems and problem sizes. On Cray XT5 system P3DFFT attains 45% efficiency in weak scaling from 128 to 65,536 computational cores. Library features include Fourier and Chebyshev transforms, Fortran and C interfaces, in- and out-of-place transforms, uneven data grids, single and double precision. P3DFFT is available as open source at http://code.google.com/p/p3dfft/. This paper discusses P3DFFT implementation and performance in a way that helps guide the user in making optimal choices for parameters of their runs.

# 1. Introduction

A large number of applications in computational sciences are written to solve partial differential equations on three-dimensional grids. Common examples are pseudospectral solvers employing Fast Fourier Transforms (FFT), as well as codes using Chebyshev transform and compact finite difference algorithms. Most of these algorithms share the feature of needing local (with respect to a processing element) access to all data in a given dimension of a 3D array at a certain stage, and then the second and third dimensions at other stages. This paper presents a software framework for such operations, and while we focus our discussion on FFTs it is naturally extended to other algorithms in this class.

The FFT in one, two or three dimensions is an efficient algorithm well known for decades [Cooley], as well as a traditional component of any serious numerical library. 3D FFT happens to be one of the most compute- and communication-intensive components in applications from a range of fields (for example in turbulence, molecular dynamics, 3D tomography and astrophysics), and for this reason there are many excellent parallel implementations in existence today. Some of the implementations are supplied by well-respected libraries and packages, whether open-source or commercial [FFTW,PESSL,NAG, Agarwal]. While these are typically highly optimized and produce good performance at moderate scales, most of them were not designed to address one increasingly important consideration, namely the limit of scalability. Most of the parallel three-dimensional (3D) FFT libraries to date use *one-dimensional (1D),* or *slab,* domain decomposition (described in Section 2) which allows scaling only up to the linear grid size. At the age of petascale platforms more and more systems typically have numbers of processing elements (PE's) far exceeding this limit for many applications. For example, cutting-edge turbulence simulations today [Donzis 2008, 2010] use $4096^3$ grids and so with 1D decomposition would only scale to 4096 PEs at most. In order to handle the enormous volume of data and compute operations such applications are run on some of the most powerful supercomputers of today, involving $O(10^4\text{-}10^5)$ PEs. The P3DFFTpackage [P3DFFT] (the subject of this paper) overcomes this scalability barrier by employing *two-dimensional* (2D), or *pencils*, domain decomposition.

While the idea of 2D decomposition applied in 3D FFT algorithm is not new, few general-purpose implementations are known to date. Wapperom et al [Wapperom] investigated a 2D decomposition applied in a FFT/Chebyshev transform and in a pseudospectral DNS turbulence code. Several interesting studies of 2D-decomposed 3D FFT on various modern supercomputer architectures have been reported recently [Takahashi 2010, Jagode, Hein, Kirker], however these codes are not available to the public. In terms of general-purpose libraries, at this time there are only a small number of offerings. An effort by an IBM team [Eleftheriou] resulted in a library of 3D-decomposed parallel FFT optimized for 3D torus architecture of IBM BG/L supercomputer. Unfortunately this software provides a very limited set of options in terms of input sizes and datatypes, and is available only as a precompiled binary for IBM BG/L. An FFT package related to LAMMPS code [Plimpton] provides array remapping tools for employing 2D decomposition and 3D FFT implementation, however this package implements only complex-to-complex transforms.

P3DFFT fills the niche by providing an open source, easy to use implementation (available from

http://code.google.com/p/p3dfft) employing 2D decomposition. By design it is portable to a large number of platforms, adaptable for diverse applications, and at the same time is built for maximum performance. P3DFFT is currently built on top of Message Passing Interface (MPI), which provides portable performance and ease of use. Its framework has been successfully applied in applications ranging from Direct Numerical Simulations of turbulence to oceanography, astrophysics, material science and chemistry [Donzis 2010, Homann 2009 and 2010, Chandy, Schumacher 2007 and 2009, Peters, Bodart, Grafke, Weidauer, Laizet, Schaeffer, informal sources]. Its performance has been rigorously studied on a torus interconnect architecture of IBM BG/L supercomputer [Chan].

The paper is structured as follows: In Section 2 we introduce the parallel 3D FFT algorithm implemented in P3DFFT as well as two most relevant decomposition schemes. In Section 3 we delve into details about P3DFFT implementation. In Section 4 we present results of some benchmark tests and analyze performance of the library comparing it with an asymptotic model. Section 5 contains a summary and conclusion.

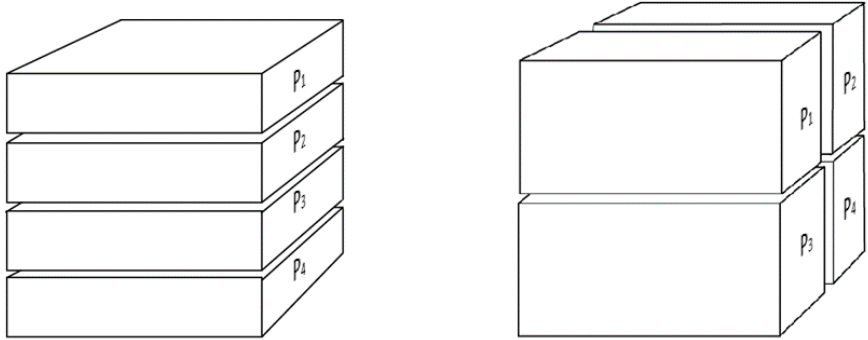

**Figure 1.** One-dimensional and two-dimensional domain decomposition. The number of data points assigned to a Processing Element $P_1$ through $P_4$ is essentially the same, but they represent different geometrical shape.

# 2. Parallel Implementation of Fourier Transforms

There are many publications devoted to parallel implementation of FFTs [Foster, Swarztrauber, Sweet and many others]. Let us begin by considering how the grid is decomposed among parallel tasks. The simplest way to distribute the data is by assigning each task one or more 2D planes (a *slab*) of the array (see Fig. 1). This is known as 1D domain decomposition because the grid is divided along one dimension. As already mentioned, this approach works well only as long as the task count $P$ does not exceed the linear grid size $N$. If taken beyond this point, 1D decomposition results in sudden loss of scalability due to load imbalance, as some tasks are without work. Two-dimensional (2D, or *pencils*) decomposition approach used in P3DFFT is the next logical step. Instead of slabs, each processor/task is responsible for a rectangular column (pencil) of the data array. The tasks are arranged in a two-dimensional virtual grid with dimensions $M_1$ x $M_2$ and this controls the dimensions of each pencil. There is some freedom in choosing the grid dimensions, as long as the product of $M_1$ and $M_2$ equals $P$, the total number of tasks. With 2D decomposition the algorithm theoretically scales up to $N^2$ tasks, and while it is somewhat more cumbersome to implement, as noted in the introduction, a number of applications running on sub-petascale and petascale platforms critically depend upon it.

Regardless of decomposition, a Fourier Transform in three dimensions is comprised of three one-dimensional (1D) FFT's in the three dimensions (X, Y and Z) in turn. When all of the data in a given dimension of the grid resides entirely in a processor's memory (i.e. it is *local*) the transform consists of a one-dimensional (1D) FFT done over multiple grid lines by every processor, which can be accomplished by a serial algorithm provided by many well-known FFT libraries and is usually a fairly fast operation. The transforms proceed independently on each processor with regard to its own assigned portion of the array. When the data are divided across processor boundaries (i.e. non-local), several approaches are possible. One approach could be to use a parallel implementation of 1D FFT that exchanges the data elements as they are needed on other processors. Such method (*distributed* transform in terminology of [Foster]) involves multiple data exchanges. An alternative is the *transpose* transform, where the array is reorganized by a single step of global transposition so that the dimension to be transformed becomes local, and then serial 1D FFT can be applied. As shown in [Foster] (see Table 1 therein), the data volume exchanged in the transpose method is approximately $log(M_1)/2$ or $log(M_2)/2$ times less than that involved in the distributed approach. As will be seen below, data volume has almost linear relation to the total execution time since communication, which comprises most of the elapsed time in many cases of interest, is dominated by the time to push the data volume through network links with limited bandwidth. Therefore the transpose approach is used in P3DFFT and throughout this paper.

Thus the parallel algorithm employed in P3DFFT (Fig. 2) has three compute stages corresponding to 1D FFT transforms in X-, Y- and Z dimensions over the entire volume of the data grid, interspersed with two parallel transposes in order to first rearrange data from X- to Y-oriented pencils, and then from Y- to Z-oriented pencils. Corresponding to the idea of the transpose approach, the 1D FFT compute stages always proceed on pencils oriented in the direction of the transform, which means the transform is local. Each transpose involves an all-to-all exchange within subgroups of tasks comprising either a row or column of the virtual 2D processor grid.

Many applications study physical problems with one dimension of non-homogeneity, for example a wall-bounded turbulent flow where two dimensions have periodic boundary conditions while the third dimension has rigid walls on both ends. For such situations it is common to use transforms other than Fourier in the third dimension, for example Chebyshev transforms, compact schemes etc The setup for these problems is essentially identical to the FFT scenario, with substitution of the appropriate third transform in place of the FFT.

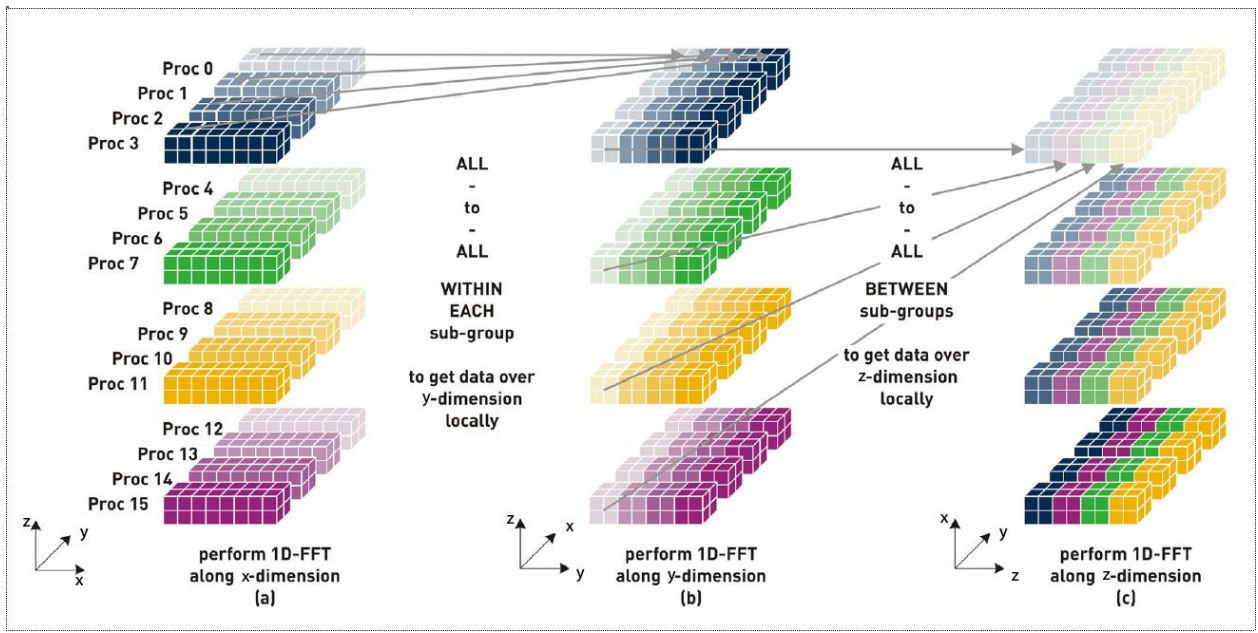

**Figure 2**. Three computation and two transpose stages of a global 3D transform with two-dimensional (2D) parallel decomposition on 16 tasks arranged in a 4x4 processor grid. Each rectangular column represents data contained locally in a processor's memory. Coordinate axes are rotated at each stage in this figure for ease of presentation. (Going from left to right) 1D FFT in X dimension for all X lines performed independently by each task is followed by an exchange/transpose among processors belonging to the same subgroup (a row in processor space). This is followed by a 1D FFT in Y followed by the second transpose among different rows, followed by the final 1D FFT in Z dimension. (Image courtesy of H.Jagode)

# 3. P3DFFT design and use

## 3.1 Library overview

P3DFFT is available as an open-source library package with an easy-to-use interface, providing implementation of 3D FFT with strong performance and scalability on a variety of platforms. Since it implements 2D decomposition it can scale up to $N^2$ cores as discussed above. 1D decomposition is also included as a special case of 2D decomposition as an option for running on moderate core counts. Both Fortran and C interfaces are supported, and comprehensive documentation is available. Also supplied in the distribution are sample programs in both Fortran and C, illustrating several usage scenarios. P3DFFT supports both in-place and out-of-place transforms in both single and double precision. It can handle any grid dimensions (i.e. not power of two) as long as underlying FFT library (ESSL or FFTW) support them. P3DFFT is capable of handling problems with uneven decomposition among processors, for example $256^3$ grid on 24 MPI tasks. Although P3DFFT works for grids of any sizes, it is designed to be most effective for large datasets. The present version 2.5 of P3DFFT implements real-to-complex (R2C forward) and complex-to-real (C2R backward) Fourier Transforms, sine/cosine (Chebyshev) transforms, as well as an empty transform which allows the user to substitute a custom transform of their own choice. P3DFFT has been tested and benchmarked on many parallel machines.

|  |  | **X**-pencil |  | **Y**-pencil |  | **Z**-pencil |  |
|---|---|---|---|---|---|---|---|
| STRIDE1 defined | $L_1$ | $N_x$ | XYZ | $N_y$ | YXZ | $N_z$, | ZYX |
|  | $L_2$ | $N_y/M_1$ |  | $(N_x+2)/(2M_1)$ |  | $N_y/M_2$ |  |
|  | $L_3$ | $N_z/M_2$ |  | $N_z/M_2$ |  | $(N_x+2)/(2M_1)$ |  |
| STRIDE1 undefined | $L_1$ | $N_x$ | XYZ | $(N_x+2)/(2M_1)$ | XYZ | $(N_x+2)/(2M_1)$ | XYZ |
|  | $L_2$ | $N_y/M_1$ |  | $N_y$ |  | $N_y/M_2$ |  |
|  | $L_3$ | $N_z/M_2$ |  | $N_z/M_2$ |  | $N_z$ |  |

**Table 1.** Local array dimensions ($L_1,L_2,L_3$) and logical memory storage order (Fortran convention) for 3D local arrays comprising X-, Y- and Z-pencils. R2C transform takes X-pencils as input and Z-pencils as output. In C2R the input/output is reversed. XYZ in this format means that the index spanning X dimension runs fastest, then Y followed by Z indices. $M_1$ and $M_2$ are dimensions of the processor grid ($M_1 \times M_2 = P$), and $N_x, N_y, N_z$ are dimensions of the 3D data grid.

## 3.2 User interface

X-pencil is defined as a local array shape such that X dimension is entirely local within a given task's memory. Y dimension is spread over the tasks in the rows of the virtual processor grid, while Z dimension is spread over columns. Similar definitions apply for Y-pencils and Z-pencils (see Table 1). P3DFFT accepts the input data array for R2C shaped as X-pencils, while output array is defined as Z-pencils. Conversely, the C2R transform expects the input array in Z-pencils while its output comes as X-pencils. Significant resources are saved by avoiding transpose back to the original distribution shape. This method is suited for a great many applications (such as convolution and differentiation algorithms) that require forward and backward transforms in sequence.

While pencil type fixes the shape of local arrays, memory storage order within the arrays can vary, depending on the pencil shape and the optional user flag STRIDE1. Memory storage dimensions are summarized in Table 1.

P3DFFT is written in Fortran90 and uses MPI for communication. While the library can be called from C as well, we follow the Fortran convention, and when talking about XYZ storage order we mean that the X dimension is the first in the order of storage, i.e. elements with different X for a given Y and Z indices are stored contiguously.

In a R2C 3D transform the first transform (in X) is a real-to-complex 1D FFT, while the following Y and Z 1D FFTs are complex-to-complex. The real-to-complex transforms are known to have complex conjugate symmetry, namely *F(x)=F\*(N-x)* which means half of the output modes are redundant due to the conjugate symmetry. In particular, in a R2C transform of size *N*, modes *N/2+2* through *N* can be reconstructed from those numbered 1 through *N/2*. Modes 1 (the average) and $N_x/2+1$ (the Nyquist frequency) have zero imaginary components. In practice this means that the output (reflected in Table 1) is represented by complex numbers in an array of dimensions *$((N_x+2)/2,N_y,N_z)$* distributed among the processors in a suitable way.

The library is called from a user's code via module interface. Internal library variables are hidden from the user while special routines are used to set or get relevant parameters such as the grid size and dimensions of pencil arrays (details of usage are described here [P3DFFT manual]). The package is easily installed using *configure* command.

## 3.3 Implementation strategy

As already mentioned, the design of P3DFFT followed the principle of portable performance. This implies that sometimes a choice needed to be made between achieving performance on a given platform and portability. A number of intermediate kernels have been created, incorporating various ideas mentioned below, which were then tested on several platforms. Some options have been found to underperform universally across the platforms, in which case the decision was clearly against them. In other cases there was no clear winner, or different variants showed up better or worse depending on the platform and/or usage scenario. In such cases we attempted to implement all of the available options and delegate the selection to the user. Thus the user has the ability to take an active part in performance tuning on their platform of choice. On the other hand, a set of reasonable defaults guarantees ease of use for those not inclined for performance timing experiments. Some of the flexibility has to do with data layout, and in this case considerations other than performance may be in place, i.e. ease of incorporating with the rest of the user's application.

As mentioned in the previous section, the compute stages of the parallel 3D FFT algorithm comprise of 1D FFT on local data. P3DFFT uses an established FFT library of user's choice (currently FFTW or ESSL) for this task. Each of the three FFTs is done as a single call, combining transforms of multiple sets of data. For transforms in the second and third dimension it is necessary to operate on data sets that are not contiguous in memory (though always local in the memory of a core/task). This can be handled in several ways. Typically FFT libraries allow both stride-1 and non-unit-stride data access patterns. Thus one approach is to let the FFT library handle the non-unit strides. Another approach is to transpose the data first to arrange them in a stride-1 format before calling FFT library. Both of these approaches are available in P3DFFT, the second of which is triggered by setting the STRIDE1 flag. Loop blocking is used with the memory transpose to optimize cache use. Relative performance advantages of these approaches are discussed in the next section. A third alternative can also be conceived, wherein FFT sub-libraries are called with input, but not the output, having unit stride (or vice versa). For example, if the input is in the form (y,z,x) where y is the fastest changing index, a call to a FFT library could transpose the input and calculate FFT in Z, and then write the output in the form (z,y,x) which is the optimal form for backward transform that follows. This approach was explored and, at least with FFTW, found to lead to inferior performance. The reasons for this are not clear at this time and may be specific to FFTW implementation. This approach is not used in P3DFFT.

As already mentioned, transposing data between pencils of different orientation between the three stages requires an all-to-all exchange within subgroups of processors. We define MPI cartesian sub-communicators ROW and COLUMN based on each task's placement in the virtual 2D processor grid. In order to transpose from X- to Y-pencils an all-to-all exchange is needed among tasks on the same rows of the processor grid. Similarly, to transpose from Y- to Z-pencils, an all-to-all exchange within columns is needed. These exchanges are implemented by a call to *MPI_Alltoall* or *MPI_Alltoallv* with ROW (COLUMN) used as the communicator argument. This relies on an optimized implementation of *MPI_Alltoall(v)* as part of MPI library on a given system. In most cases *MPI_Alltoall(v)* can be expected to work faster than an equivalent collection of point-to-point send/receive calls. This was confirmed in numerical experiments on several platforms. (The above is valid for blocking operations. Using non-blocking point-to-point sends/receives, or PGAS languages such as Co-Array Fortran, allows overlapping communication and computation. This is an interesting alternative, however it is not always a portable choice. Studies are ongoing, and future releases of the package may include a PGAS or overlap option.)

The parallel transposes involve packing/unpacking data into/out of send/receive buffers. Essentially this procedure involves a memory copy. In addition, when STRIDE1 option is defined the packing/unpacking may involve a local memory transpose, i.e. the copy operation does not follow a stride-1 pattern. We use loop blocking to minimize cache misses. While MPI standard provides a mechanism for such pack/unpack operations through MPI datatypes, performance is dependent on the particular MPI implementation. In experiments to date no measurable gain in performance was found for a version implemented with MPI Datatypes. Therefore we chose to perform the packing/unpacking explicitly.

## 3.4 Load balancing

The algorithm lends itself to a symmetric work distribution, and P3DFFT attempts to divide the work evenly among the tasks (this also includes communication). In cases of uneven dimensions for local grid (for example when using 128x128x128 data grid on 6 CPUs) load imbalance is naturally present. In practice there is a small load imbalance even in the case when the processor grid evenly divides the data grid due to the nature of R2C and C2R transforms, since as mentioned earlier the dimensions of the output array is *($(N_x+2)/2,N_y,N_z$)*. This leads to a slight load imbalance, even if the initial problem is evenly decomposed by the processor grid. Since the number of elements exchanged among tasks may not be the same, the *MPI_Alltoallv* call is used for exchanges as it allows for more flexibility in defining data than *MPI_Alltoall*. The mentioned load imbalance is quite small for a reasonably large data grid volume, and in practice does not cause performance problems. This implementation is therefore well suited for evenly as well as unevenly distributed data grids.

While *MPI_Alltoallv* is the simplest solution, it is not necessarily the best one in some cases. In particular, it has been reported [Schulz] that implementation of *MPI_Alltoallv* on Cray XT platforms is inferior in performance to *MPI_Alltoall* for comparable (evenly divided) buffers. For this reason P3DFFT provides an option (USEEVEN) for reverting to *MPI_Alltoall* by simply padding send buffers with a few extra elements in order to make the exchanges even. When load imbalance is small these extra elements do not substantially add to the volume of data exchanged and the end result is better that using MPI_Alltoallv on these systems, as shown below in the results section.

# 4. Results and Discussion

## 4.1 Benchmark tests description

In this section we describe benchmark results obtained by using one of the sample programs (test_sine) supplied with the P3DFFT package. This program initializes a 3D array of specified dimensions, and performs a forward and backward 3D FFT, then checks to make sure the data is the same (apart from a scale factor) as the initial array. The forward and backward FFT can be repeated in a timed loop for a specified number of iterations. The program reports time results averaged over the iterations count.

P3DFFT has been tested on many platforms, and in this paper we discuss results obtained on Cray XT5 partition of **Jaguar** at NCCS/ORNL, **Kraken** at NICS/University of Tennessee, and the **Ranger** system at TACC/University of Texas at Austin. Jaguar is a Cray XT5 computer with 2.6 GHZ AMD Opteron processors, 12 cores per node (2 sockets with 6 cores per socket), 16 GB of memory per node, and Cray SeaStar2 3D torus interconnect [Jaguar], with a total of 224,256 compute cores. GNU compilers were used in these tests. Kraken is a similar but somewhat smaller system at NICS/UT [Kraken]. Ranger is a Sun/AMD platform with 2.3GHz AMD Opteron processors, 16 cores per node (4 quad-core), 32 GB of memory per node, and a Closs interconnect (InfiniBand switch), with a total of 62,976 compute cores [Ranger]. PGI compilers were used in our tests on Ranger.

## 4.2 Achieving optimal performance with P3DFFT

As mentioned earlier, P3DFFT was designed with reasonable defaults for portable performance, as well as several user-controlled settings that can be adjusted for maximum performance. These settings include the STRIDE1 option, the USEEVEN option, and a suitable choice of the aspect ratio of the virtual processor grid. We are going to discuss these in turn.

1. **STRIDE1 option**

Recall that STRIDE1 option controls whether the local memory transpose is performed in P3DFFT or delegated to the FFT sub-library. There are pros and cons for each choice. On the one hand, delegating the transpose to the FFT library allows it to optimize cache flow in combination with the FFT itself. On the other hand, in this case the non-stride-1 access is needed both on input and output of the FFT. If the data is transposed and written to a temporary buffer, as in P3DFFT, separately from FFT, the out-of-cache access pattern is needed only in one of either input or output. The cost of this is an extra memory write/read, which is done in small chunks, guaranteeing that the data reside in cache. In general there is no clear winner between these two approaches, as many factors contribute (size of transforms, cache architecture of a given machine, sophistication of a given FFT implementation's handling of non-stride-1 data). In many tests performance of the two approaches has been found quite close. In practice this choice is often tied up with how the data structures are set up in the application that uses P3DFFT (Fourier space arrays are ordered ZYX as opposed to XYZ by default, per Table 1).

2. **USEEVEN option**

In the previous section we mentioned potential advantage of using *MPI_Alltoall* call to transpose arrays even if the buffers are of different sizes for all the tasks, by padding the buffers. P3DFFT provides USEEVEN option to use this implementation, whereas the default is *MPI_Alltoallv*. The effect of using this option is quite noticeable on Cray XT5 (see Figure 4 below), due to an abnormal difference between *MPI_Alltoall* and *MPI_Alltoallv* performance mentioned earlier [Schulz]. The USEEVEN option is therefore strongly recommended on Cray XTs but is not necessary on other platforms.

3. **Processor grid dimensions and task placement**

Below we consider the main performance bottleneck for P3DFFT, namely the two transposes involving all-to-all communication.

Let us assume that the messages are large enough so that the bandwidth of the network is the main bottleneck for inter-processor communication, rather than the latency. The all-to-all communication pattern in 3D FFT depends mainly on two factors: bisection bandwidth of the network and the degree of contention between different messages in utilization of the network links. The latter contention is a function of a given network topology, as well as the efficiency of the all-to-all algorithm and its implementation in the system. Since current generation of P3DFFT uses MPI_Alltoall(v) (in order to be portable yet to use each system's optimized all-to-all), in this paper we do not attempt to model network contention and related effects but focus on network throughput defined by the bisection bandwidth as a rough upper limit estimate. Some in-depth modeling work for 3D torus architecture of the IBM BG/L network, taking into account contention effects, can be found in [Chan, Kumar, Pekurovsky].

First consider a 3D grid decomposed in one dimension. In the 3D FFT problem the number of floating point operations in computation phases is $O(N^3 log(N))$ while the data volume for all-to-all communication is $O(N^3)$. As the number of tasks (or cores) P is increased, the computation time scales as $O(1/P)$ since the work is nearly evenly distributed over P tasks. An all-to-all exchange involves each task sending and receiving P-1 messages of size $N^3/P^2$, so the total volume of data is approximately $m*N^3$ (here m is the size of each array element in bytes, and $N^3$ can be substituted with the product of $N_x * N_y * N_z$ for non-cubic grids). By definition, in the all-to-all exchange every second message will cross the network bisection, and therefore the time for such transfer is approximated by

$$T_{network} = m\ c\ N^3 / (2\ \sigma_{bi}(P)), \qquad (1)$$

where $\sigma_{bi}(P)$ is bisection bandwidth of the network portion containing P tasks, and c is a constant containing network contention and other effects.

As already mentioned, in a 2D decomposition there are two all-to-all exchanges involved, each performed within a ROW or COLUMN sub-communicator. Although they exchange the same volume of data, they have rather different properties due to the patterns of topological placement of tasks on the network. The ROW sub-communicators normally consist of tasks on adjacent nodes (or even within a single node). These nodes are going to end up somewhere close in their location on the network graph. However each COLUMN sub-communicator will span tasks on nodes that are spread out across the network. Since performance characteristics of all-to-all exchange substantially depend on how far the nodes are from each other on the network graph, typically the row- and column exchanges have very different properties, so there is an asymmetry in their parameters. For this reason, when choosing the processor grid dimensions, a square grid is often not optimal, even if it may seem counter-intuitive. The optimal choice depends on specifics of the architecture of a given platform. The dimensions should satisfy

$$M_1 * M_2 = P,\ M_1 <= (N_x/2, N_y)\ and\ M_2 <= (N_y, N_z) \qquad (2)$$

Assuming there is no problem with injecting many messages into the network, often the best dimensions are such that the ROW exchange is done entirely within the node, or among a few neighboring nodes. (This means that $M_1 << M_2$.) In such scenario, the ROW exchange will be defined by memory bandwidth on the node and quite cheap in comparison with the COLUMN exchange. In contrast, in the case where $M_1$ and $M_2$ are of the same order, both ROW and COLUMN exchanges involve network traffic across nodes, resulting in a higher total cost.

**Therefore choosing $M_1$ such that the ROW transpose occurs within one node, if feasible, appears to be a good strategy.**

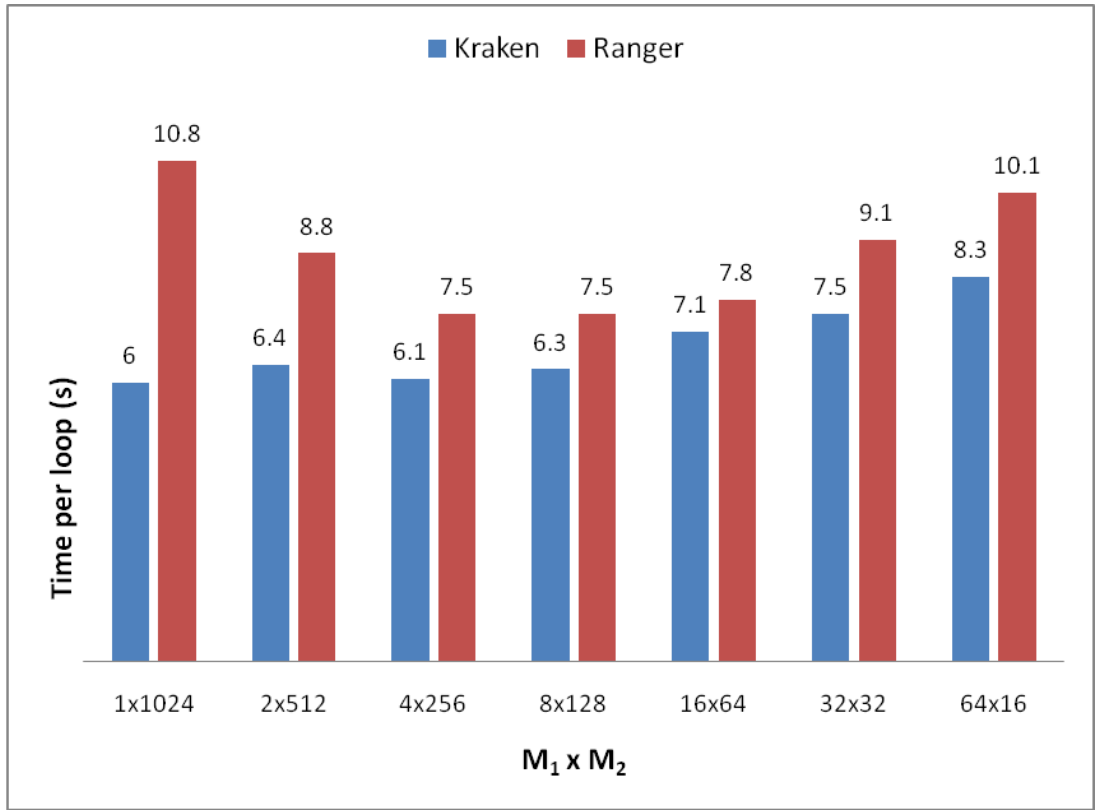

**Figure 3.** Performance dependence on processor grid aspect ratio $M_1 \times M_2$. Each bar's height shows the time to complete one forward and one backward 3D FFT on a $2048^3$ grid using 1024 cores on Cray XT5 (Kraken) and Sun/AMD (Ranger).

In Figure 3 we study the dependence of performance on processor grid dimensions. Data shown confirm the benefit of choosing $M_1$ such that the ROW transpose occurs within a node. We see that performance of P3DFFT can depend noticeably on the choice of processor grid aspect ratio. In this example ($2048^3$ grid on 1024 cores) the time to solution rises as $M_1$ crosses the threshold equal to the number of cores per node (in this case 12 for Kraken, 16 for Ranger). Note that the square grid 32x32, which may seem intuitively the best one, does not yield the optimal performance. The reason for reduced performance on Ranger with low $M_1$ is not clear at this time, possibly related to more messages leaving each node in the COLUMN exchange when $M_1$ is low.

It is interesting to note that in our tests on Cray XT5 (Kraken) at high core counts (see Fig. 4) the situation is somewhat different, and performance favors more square-shaped processor grids.

One hypothesis explaining this is a limitation on the number of messages that the Cray's SeaStar interconnect can handle at a given time.

We conclude that the choice of optimal processor grid dimensions appears to be architecture- and problem-specific. The value of $M_1$ being close to the number of cores on the node is a good starting point in many cases, however the user is advised to carry out a few test runs with higher and lower values for each case.

Assuming the above is the case, Eq. 1 is still a good approximation for total communication time, since now the COLUMN exchange follows the same pattern as the all-to-all exchange in 1D decomposition. Therefore total execution time for 3D FFT can be approximated by the following:

$$T_{FFT} = N^3 [2.5 \log(N)/(P\,F) + b\,m/(P\sigma_{mem}) + c\,m/(2\,\sigma_{bi}(P))]. \tag{3}$$

Here the parameter *b* contains the number of memory accesses per data element, both in FFT operations and in all the local and non-local transposition steps, while $\sigma_{mem}$ is memory bandwidth per task on a node . Parameter *F* reflects the floating point operations per second in FFT. The values of coefficients *F, b* and *c* vary depending on the system hardware, as well as implementation of MPI and FFT library. The above is a model that will be used in the following sections to fit the measured timing data. For cases when $M_1$ is larger than the number of cores on the nodes, both ROW and COLUMN exchanges involve network traffic, but even in the worst case both of them are limited by bisection bandwidth and therefore the Eq. 3 should still hold.

In our studies to date, the default placement of tasks on the physical network (i.e. cores on a node are populated with contiguous task IDs) has proven to be optimal, compared with several other task mapping algorithms, when performing 3D FFT on cubic grids. Essentially this is due to the fact that all-to-all operations are asymmetric and not sensitive to task placement. However, for non-cubic grids the situation is different. An interesting study was reported by an ANL group [Chan] where up to 48% improvement in performance was found due to optimized task placement based on the network topology when using P3DFFT with non-cubic domains on IBM BG/L. While further work here is needed, in this paper we focus on results obtained when using simple contiguous task placement and study cubic grids only.

### 4.3 Benchmark Results and Discussion

Figures 4 and 5 show strong scaling of the P3DFFT forward-and-backward 3D FFT test of size $4096^3$ using double precision on Cray XT5 (Kraken). Since we have seen that processor grid aspect ratio has some effect on performance, only the best $M_1$ x $M_2$ combination is taken as data point for each core count. It is clear from these figures that communication time is a substantial part of the total execution time.

Recall that this platform has 3D torus interconnect, and therefore bisection bandwidth scales asymptotically as $O(P^{2/3})$. Therefore Eq. 3 becomes:

$$T_{FFT} = a/P + d/P^{2/3} \qquad (4)$$

A least-squares fit to the data to a curve defined by Eq. 4 produces an excellent match. Taking the value of coefficient of the $P^{2/3}$ term obtained from the fit, assuming that ½ of all messages will need to pass through the network bisection, and considering that there are two transposes, we obtain a value of 212 GB/s for effective bandwidth for the network portion containing 65,536 cores (or 5,462 nodes). It is not easy to contrast this with a predicted value since the placement of tasks on the torus is unknown. If we assume 15x16x24 partition with wraparound torus links in the second and third dimension, and peak bandwidth of 9.6 GB/s per link, the expected bisection bandwidth is 16*24*9.6 GB/s = 3,686 GB/s, which implies about 6% efficiency. This number parameterizes network contention and any other loss of efficiency in the network. This is a rough estimate and should be taken with caution due to many unknowns, such as how the nodes were grouped together, and whether other users' traffic affected the network throughput. However, the results of this analysis are in agreement with the assumption that communication time scaling is inversely proportional to bisection bandwidth.

P3DFFT performs and scales reasonably well not only for large transforms, but also for medium grid sizes. Figures 6-8 present benchmark results for tests with linear grid sizes 2048, 1024 and 512.

A weak scaling study is presented in Figure 9. We compare timings for 5 cases, involving 16, 128, 1024, 8,192 and 65,536 compute cores, with corresponding grid sizes $512^3$, $1024^3$, $2048^3$, $4096^3$ and $8192^3$. Computational intensity of 3D FFT is $O(N^3 log(N))$. Therefore the weak scaling can be approximately defined by increasing the core count 8 times with each two-fold increase in grid size, and including a factor of $log(N)$ in the efficiency. With this definition P3DFFT achieves parallel efficiency of 45% over the range of core counts from 128 to 65,536.

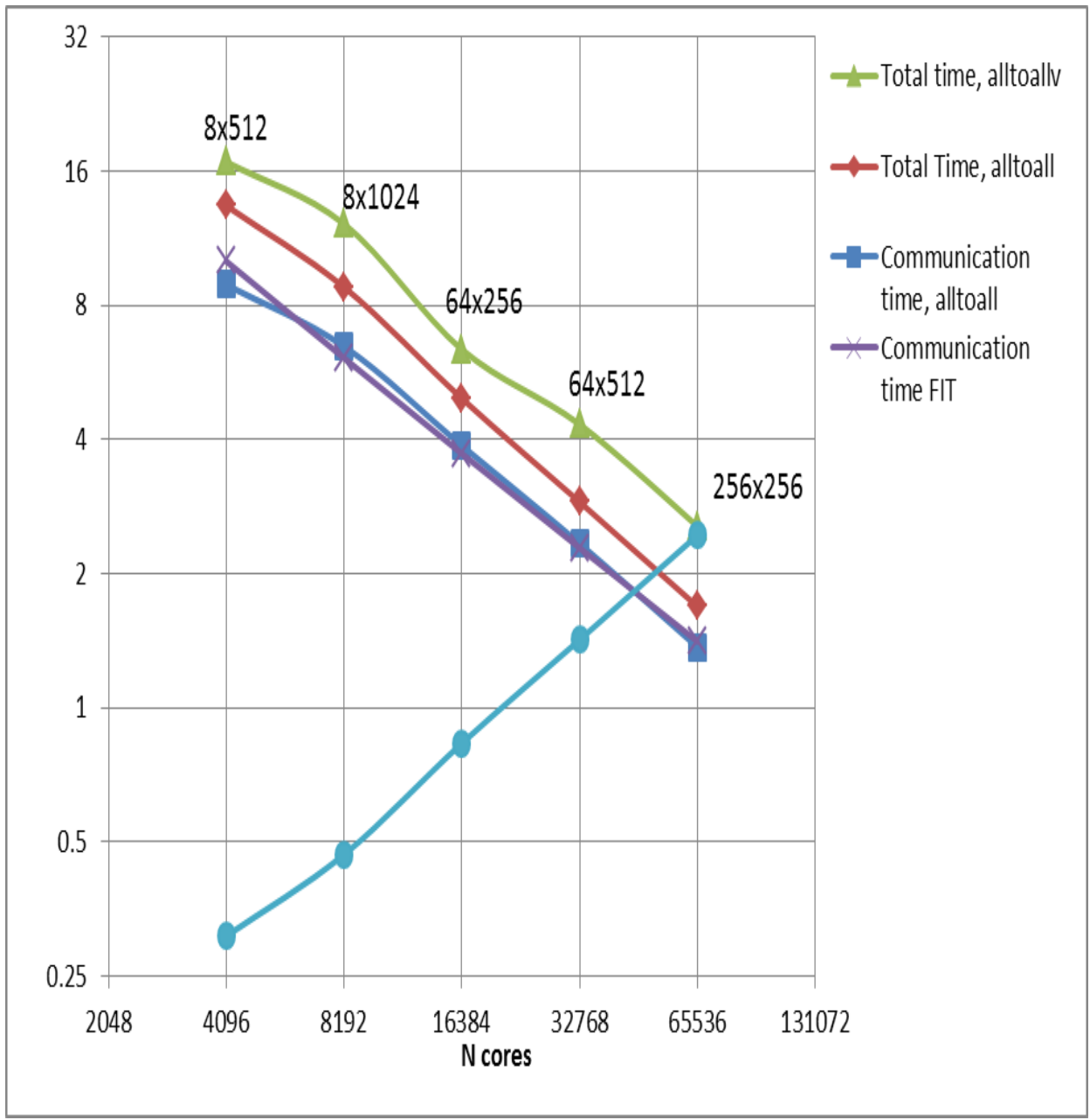

**Figure 4.** Strong scaling of P3DFFTbenchmark for a $4096^3$ grid in double precision on Cray XT5. This log-log plot compares performance of versions implemented with *MPI_Alltoall* (USEEVEN option, red diamonds) and *MPI_Alltoallv* (default, green triangles) for parallel transposes. Both versions use STRIDE1 option, however results without STRIDE1 are very close to these numbers. Also shown is communication time corresponding to *MPI_Alltoall* version (blue squares) and a calculated fit to this data of the function $a/P + b/(P^{2/3})$ (magenta crosses). Numbers next to each data point show the processor grid dimensions used in obtaining the timing, corresponding to best result among several geometries explored for each case. Both time in seconds (for a forward/backward transform pair) and TeraFlops (achieved number of floating point operations per second, in $10^{12}$) are plotted. (Tflops numbers are based on best results with alltoall).

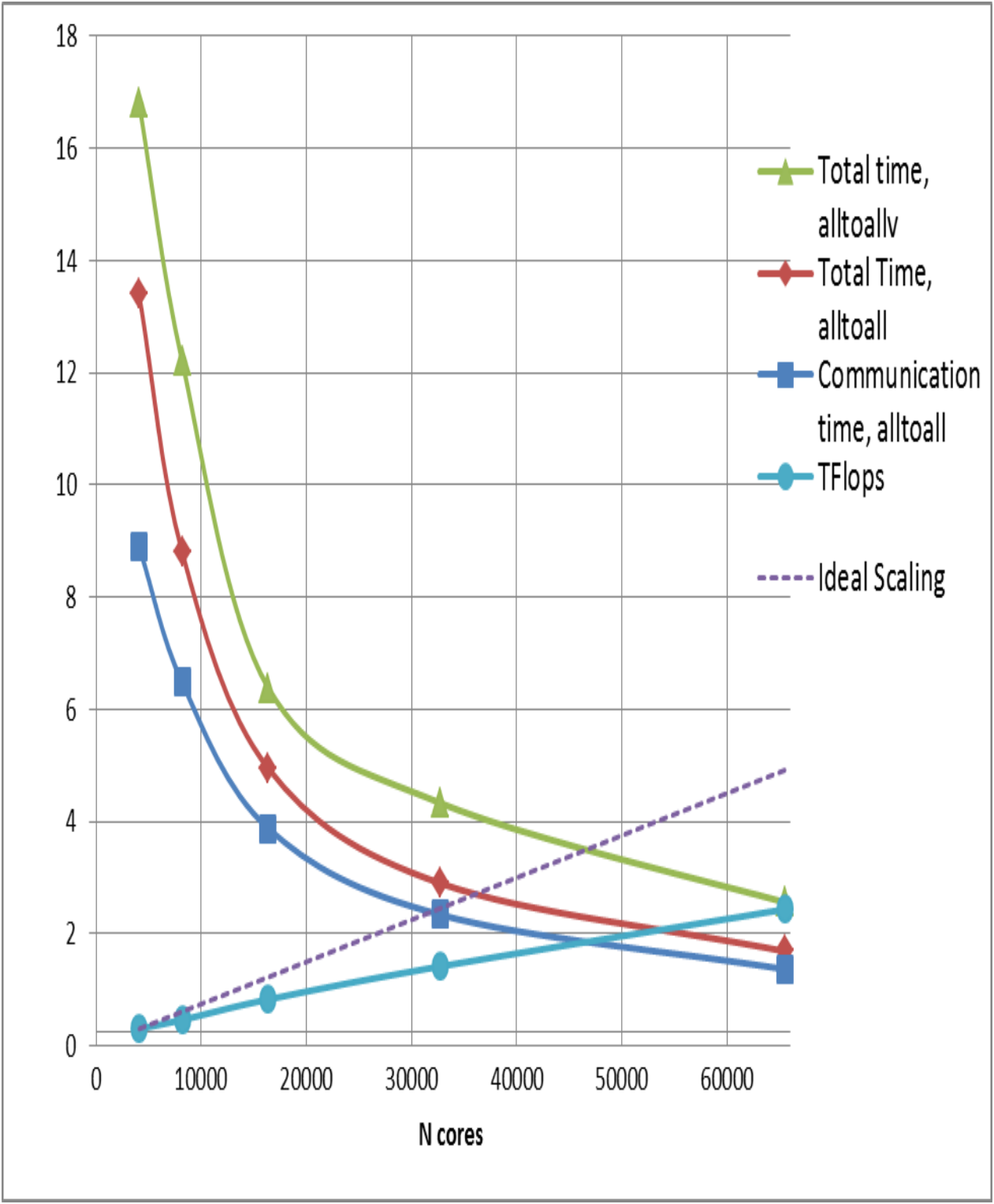

**Figure 5.** Same data as in Figure 4 plotted with linear (instead of logarithmic) axes.

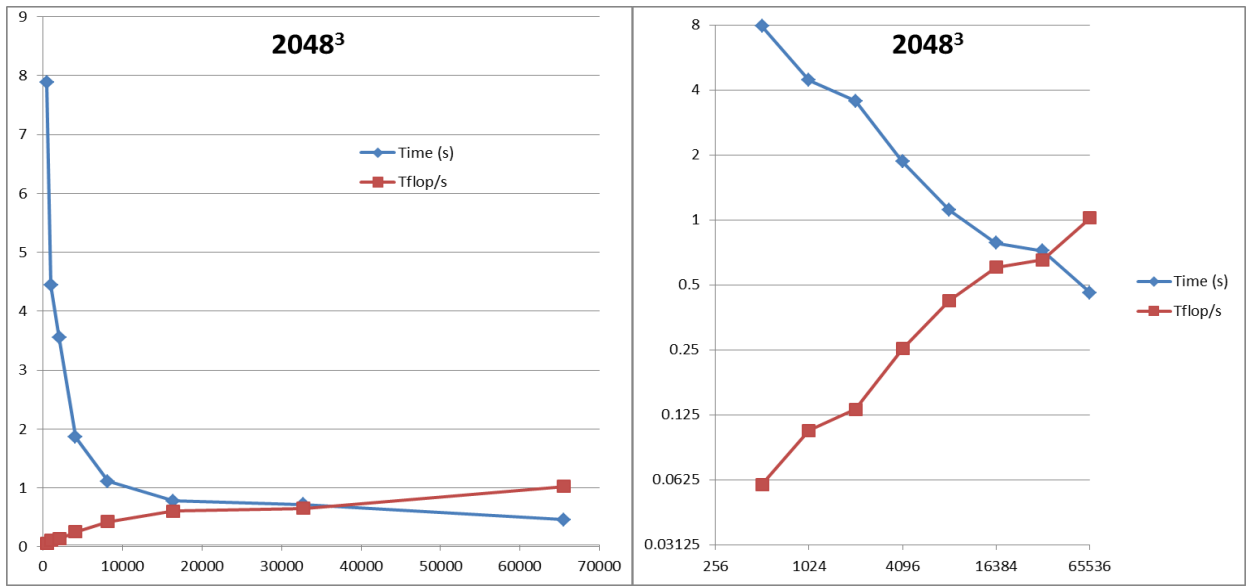

**Figure 6.** Performance of P3DFFT on a $2048^3$ transform, running on Cray XT5. Both time in seconds (for a forward/backward transform pair) and number of floating points per second (in $10^{12}$, or TeraFlops) are plotted, and both linear and log-log plots are presented.

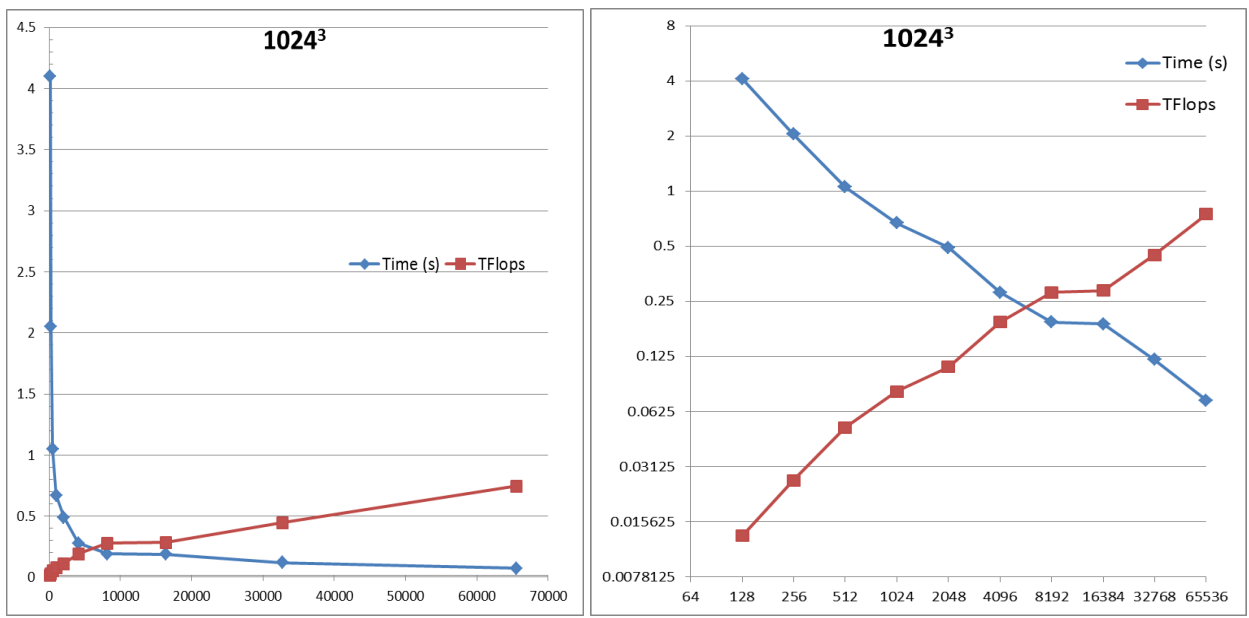

**Figure 7.** Performance of P3DFFT on a $1024^3$ transform, running on Cray XT5. Both time in seconds (for a forward/backward transform pair) and number of floating points per second (in $10^{12}$, or TeraFlops) are plotted, and both linear and log-log plots are presented.

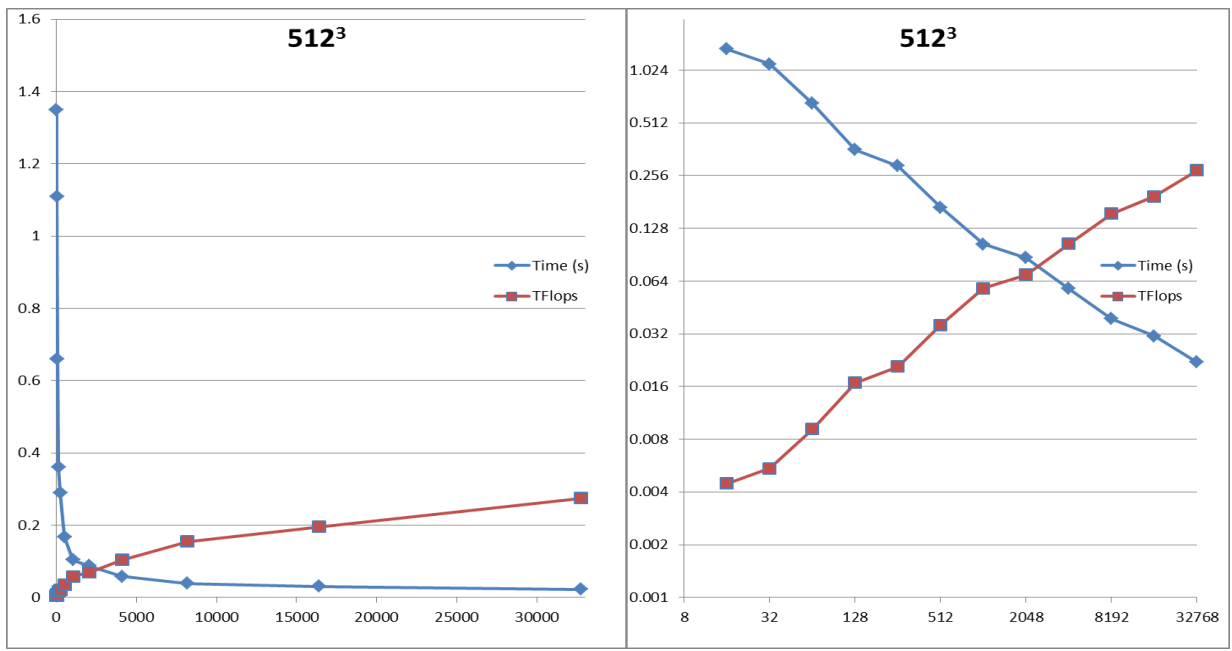

**Figure 8.** Performance of P3DFFT on a 512³ transform, running on Cray XT5. Both time in seconds (for a forward/backward transform pair) and number of floating points per second (in $10^{12}$, or TeraFlops) are plotted, and both linear and log-log plots are presented.

Finally we touch on advantages and disadvantages of using 2D decomposition versus 1D decomposition in 3D FFT at large and moderate scales. Both 1D and 2D decompositions are implemented by P3DFFT (1D is simply a special case of 2D with virtual processor grid geometry *1 x P*). While 2D version is free from the scaling limitation of slabs distribution, at moderate scales (namely *P <= N*) the 1D decomposition may be better, and our study in Figure 10 confirms this. The main reason for the difference is that in 1D case there is only one transpose instead of two. The difference gets smaller as core count P is increased, becoming negligible at *P=N*. As *P* goes beyond *N*, it is clear that the 1D version cannot scale, however the 2D version continues to scale. Therefore the main reason for using the 2D approach is maintaining scalability beyond the *P=N* point.

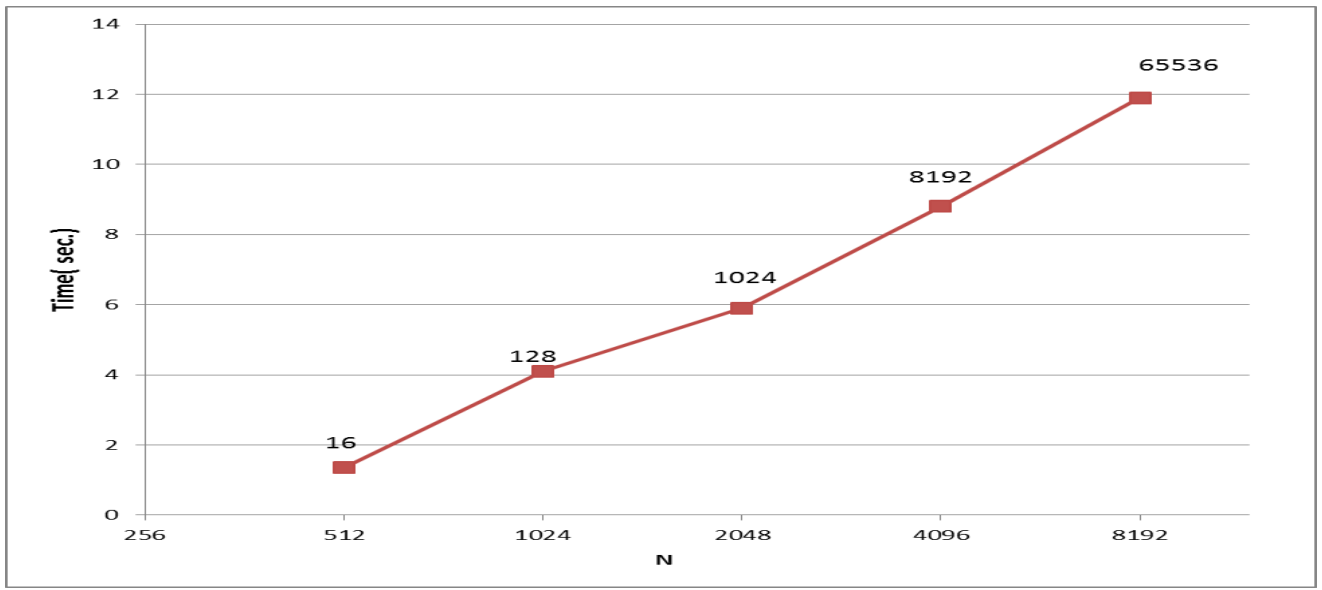

**Figure 9.** Weak scaling of P3DFFT test on Cray XT5. The horizontal axis shows linear data grid size. The numbers next to data points indicate the core count. This is a log-linear plot.

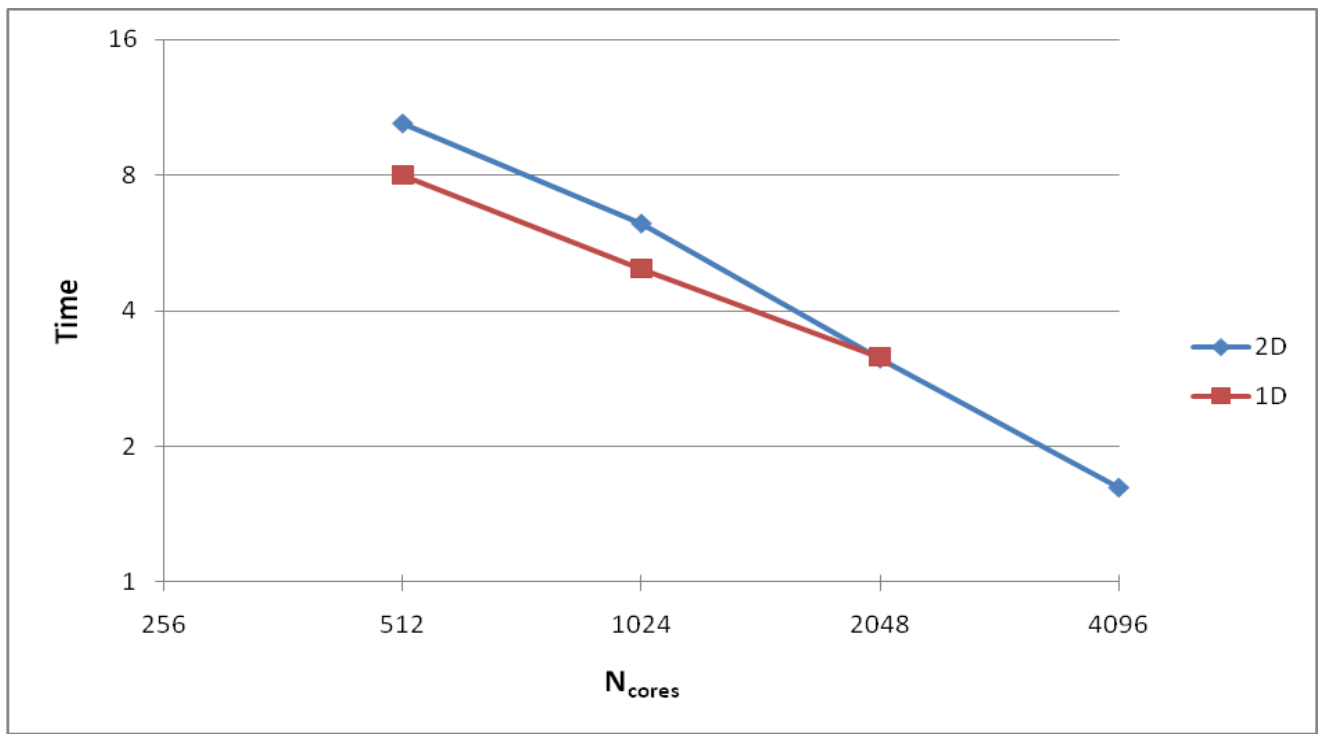

**Figure 10.** Scaling of 1D and 2D vesions of the test FFT program. This study was done on Cray XT5 (Kraken) using $2048^3$ grid in double precision, with GNU compilers. There is no 1D data at 4096 cores since the number of tasks exceeds the number of slabs available (in this case 2048).

# 5. Conclusion and related work

In this paper we have presented P3DFFT, a software package for fast three-dimensional Fourier transforms and related algorithms on parallel computers. The algorithm is implemented using two-dimensional decomposition, which significantly increases the range of scaling compared with one-dimensional decomposition used in many other implementations. The design of P3DFFT followed the goal of portability and performance and is based on a thorough set of experiments.

The main factors affecting performance were analyzed in this paper in a way that helps guide the user in choosing optimal parameters for their run. The paper also contains some benchmark data which agree well with an asymptotic model based on the network bisection bandwidth for the studied architectures. The reader is given a realistic level of expected performance achievable with P3DFFT at high scales.

On Cray XT5 system P3DFFT achieves a weak scaling efficiency of 45% when scaling from 128 to 65,536 compute cores. Loss of efficiency is attributed to the difficulty of maintaining scalable sustained bisection bandwidth on systems with core counts reaching into tens and hundreds of thousands. While this tendency is likely to be exacerbated in the future, at present this is a reasonable level of scalability to enable many critical computations at large core counts. P3DFFT has already found application in many scientific projects [Donzis 2008 and 2010, Homann 2009 and 2010, Chandy, Schumacher 2007 and 2009, Peters, Bodart, Grafke, Weidauer, Laizet, Schaeffer].

Studies involving Unified Parallel C (UPC) implementation of 3D FFT have been reported in a number of papers including [Bell, Nishtala]. The authors report seeing an advantage in performance with UPC over MPI (using *MPI_isend* and *MPI_irecv*) at medium scales, which they attribute to overlap of communication with computation. By using communication protocols such as GASNET that are closer to the low-level network fabric than MPI, a higher efficiency of overlap is achieved. Another work with communication overlap in 3D FFT has been reported in [Kandalla], where a gain in performance is shown due to overlap of communication and computation using MPI_Put calls. In general the ability to achieve overlap is highly dependent on the capabilities of the hardware and software of a given system. It may be also noted that while this avenue deserves careful attention and study, measurements on production high-end systems (for example Cray XT) show that on the order of 80% of total time is spent in communication at high core counts with 3D FFT, which unfortunately limits the gains achievable with overlap of communication and computation in this case.

Several teams have studied hybrid MPI/OpenMP implementation of pseudospectral CFD codes employing 3D FFTs [Mininni, Gorobets, Takahashi 2006, Itakura]. Using hybrid code may potentially help performance somewhat by aggregating messages in all-to-all exchanges, although it is not likely to relieve the fundamental problem of limited bisection bandwidth. The downside is possible performance loss due to OpenMP overhead and false sharing, as well as increased level of programming difficulty. While questions remain about whether the hybrid approach provides a decisive advantage over pure MPI, it is being investigated and may be implemented in later versions of P3DFFT package.

As mentioned earlier, P3DFFT implements several transform types in the third dimension, including Chebyshev and empty transform, for those cases where the third dimension requires special treatment with respect to the first two. Future evolution of the package will follow the need of its users, and may include increased flexibility of array layout, as well as a versatile collection of isolated array transpose calls. This will broaden the range of applicability of P3DFFT in scientific computing.

# Acknowledgements

This work was sponsored by National Science Foundation (NSF) grant OCI-085-684. The author acknowledges use of NSF computer resources at NICS/ORNL, and the TACC/University of Texas Ranger system, under a Teragrid startup allocation award, as well as Cray XT5 (Jaguar) system at NCCS/ORNL under a DOE INCITE award "A Petascale Study of Turbulent Mixing in Non-Stratified and Stratified Flows". The author would also like to thank P.K.Yeung, D. Donzis, R. Schulz, and G. Brethouwer for helpful discussions.

# References

R.C.Agarwal, F.G.Gustavson, M.Zubair (1994), *An efficient parallel algorithm for the 3-D FFT NAS parallel benchmark*. In: Proceedings of the Scalable High-Performance Computing Conference, pp. 129–133

C. Bell, D. Bonachea, R. Nishtala, K. Yelick (2006), *Optimizing bandwidth limited problems using one-sided communication and overlap*, Parallel and Distributed Processing Symposium (IPDPS 2006).

J. Bodart (2009), *Large scale simulation of turbulence using a hybrid spectral/finite difference solve*r, in Parallel Computational Fluid Dynamics: Recent Advances and Future Directions, pp. 473-482.

A. Chan, P.Balaji, W.Gropp, R.Thakur (2008), *Communication analysis of parallel 3D FFT for flat cartesian meshes on large Blue Gene systems,* HiPC'08 Proceedings of the 15th international conference on High Performance Computing, Bangalore, India, pp. 350-364

A.J.Chandy, S.H.Frankel (2009), *Regularization-based sub-grid scale (SGS)models for large eddy simulations (LES) of high-Re decaying isotropic turbulence*, Journal of Turbulence, Volume 10, N 25, p.1.

J.W.Cooley, J.W. Tukey (1965), *An algorithm for the machine calculation of complex Fourier series*, Math. Comput. 19, 297–301

D. A. Donzis, P. K. Yeung, D. Pekurovsky (2008), *Turbulence simulations on $O(10^4)$ Processors*, In: TeraGrid'08 Conference, Las Vegas, NV.

D. A. Donzis, P. K. Yeung (2010), *Resolution effects and scaling in numerical simulations of passive scalar mixing in turbulence*, Physica D 239, 1278–1287.

M. Eleftheriou, B. Fitch, A. Rayshubskiy, T.J.C. Ward, and R.S. Germain(2005), *Performance measurements of the 3D FFT on the Blue Gene/L supercomputer*, Euro-Par 2005 Parallel Processing: 11th International Euro-Par Conference, Lisbon, Portugal, Volume 3648 of Lecture Notes in Computer Science, edited by J.C. Cunha and P.D. Medeiros. Springer-Verlag, 2005. Pp. 795–803

I. Foster, P. Worley (1997), *Parallel algorithms for the spectral transform method*, SIAM journal on scientific computing, Volume:18, Issue 3, Page 806.

M.Frigo, S.G.Johnson (2005), *The design and implementation of FFTW3*. Proc. IEEE 93, 216–231

A. Gorobets, F.X. Trias, R. Borrell, O. Lehmkuhl, A. Oliva (2011), *Hybrid MPI + OpenMP parallelization of an FFT-based 3D Poisson solver with one periodic direction*, Computers & Fluids, Volume 49, Pages 101–109.

G. Grafke (2008), *Numerical simulations of possible finite time singularities in the incompressible Euler equations: Comparison of numerical methods*, Physica D: Nonlinear Phenomena, Volume 237, Pages 1932-1936.

J. Hein, H. Jagode, U. Sigrist, A. Simpson, A. Trew (2008), *Parallel 3D-FFTs for multicore nodes on a mesh communication network*, In: CUG Proceedings, 2008

H.Homan et al. (2009), *Bridging from Eulerian to Lagrangian statistics in 3D hydro- and magnetohydrodynamic turbulent flows*, New J. Phys. 11 073020.

H. Homann et al. (2010), *DNS of Finite-Size Particles in Turbulent Flows*, in "John von Neumann Institute for Computing NIC Symposium 2010 Proceedings, 24 - 25 February 2010 : Juelich, Germany", pp. 357-364.

K. Itakura, A. Uno, M. Yokokawa, T. Ishihara, Y. Kaneda (2004), *Scalability of hybrid programming*

*for a CFD code on the Earth Simulator,* Parallel Computing, Volume 30, Pages 1329–1343.

H. Jagode (2008), *Custom Assignment of MPI Ranks for Parallel Multi-dimensional FFTs: Evaluation of BG/P versus BG/L*, Parallel and Distributed Processing with Applications, ISPA '08, 271 – 283

Jaguar, http://www.nccs.gov/computing-resources/jaguar/

K. Kandalla, H. Subramoni, K. Tomko, D. Pekurovsky, S. Sur, D. Panda (2011), *High-performance and scalable non-blocking all-to-all with collective offload on Infiniband clusters: a study with parallel 3D FFT*, Computer Science – Research and Development, Volume 26, Issue 3, Pages 237-246 (ISC'11)

Kraken, http://www.nics.tennessee.edu/?q=node/38

I. Kirker (2009), *Demanding Parallel FFTs*: Slabs & Rods, Master's Thesis, EPCC, http://www.epcc.ed.ac.uk/wp-content/uploads/2009/01/Ian_Kirker.pdf

S. Kumar, Y. Sabharwal, R. Garg, P. Heidelberger (2008), *Optimization of All-to-All Communication on the Blue Gene/L Supercomputer*, 37th International Conference on Parallel Processing, 2008. ICPP '08, Pages 320 – 329.

S. Laizet et al. (2010), *A numerical strategy to combine high-order schemes, complex geometry and parallel computing for high resolution DNS of fractal generated turbulence*, Computers & Fluids, Volume 39, Issue 3, Pages 471-484.

Pablo D. Mininni, Duane L. Rosenberg, Raghu Reddy, Annick Pouquet (2011), *A hybrid MPI-OpenMP scheme for scalable parallel pseudospectral computations for fluid turbulence*, Parallel Computing, Volume 37, Pages 316-326.

NAG (Numerical Algorithms Group) library, http://www.nag.com

R. Nishtala, P. Hargrove, D. Bonachea, K. Yelick (2009), *Scaling communication-intensive applications on BlueGene/P using one-sided communication and overlap*, IEEE International Symposium on Parallel & Distributed Processing (IPDPS 2009), Pages 1-12.

D. Pekurovsky, P.K. Yeung ,D. Donzis, S. Kumar, W. Pfeiffer, G. Chukkapalli (2006), *Scalability of a pseudospectral DNS turbulence code with 2D domain decomposition on Power4+/Federation and Blue Gene systems,* Proc. ScicomP12, Boulder,Colorado, http://www.spscicomp.org/ScicomP12/www.cisl.ucar.edu/info/scicomp/Presentations%20CD/7-20_Thursday/Session%201/Pekurovsky.pdf

PESSL (Parallel Engineering and Scientific Subroutine Library)
http://publib.boulder.ibm.com/infocenter/clresctr/vxrx/topic/com.ibm.cluster.essl.doc/pessl_aix33/am60130515.html

N. Peters et al.(2010), *Geometrical Properties of Small Scale Turbulence*, In: "John von Neumann Institute for Computing NIC Symposium 2010 Proceedings, 24 - 25 February 2010 : Juelich, Germany", pp.365-371.

S. Plimpton, *Parallel FFT Package*, http://www.sandia.gov/~sjplimp/docs/fft/README.html

P3DFFT Manual, http://code.google.com/p/p3dfft/downloads/detail?name=P3DFFT_USER_GUIDE_2.4.pdf

Ranger, http://services.tacc.utexas.edu/index.php/ranger-user-guide

P. Schaeffer et al.(2010), *Testing of Model Equations for the Mean Dissipation using Kolmogorov Flows*, Flow, Turbulence and Combustion.

J. Schumacher, M. Putz (2007), *Turbulence in Laterally Extended Systems*, Parallel Computing: Architectures, Algorithms and Applications, C. Bischof, M. Bucker, P. Gibbon, G.R. Joubert, T. Lippert, B. Mohr, F. Peters (Eds.), In: John von Neumann Institute for Computing, Julich, NIC Series, Vol. 38, ISBN 978-3-9810843-4-4, pp. 585-592.

J. Schumacher (2009), *Lagrangian studies in convective turbulence*, Phys.Rev. E 79, 056301.

R. Sweet, W. Briggs, S. Oliveira, J. Porsche, T. Turnbull (1991), *FFTs and three-dimensional Poisson solvers for hypercubes*, Parallel Computing, Volume 17 (1991), Pages 121-131.

P. Swarztrauber (1987), *Multiprocessor FFTs*, Parallel Computing, Volume 5, Issues 1-2, Pages 197-210.

D. Takahashi (2006), *A Hybrid MPI/OpenMP Implementation of a Parallel 3-D FFT on SMP Clusters*, Parallel Processing and Applied Mathematics, Lecture Notes in Computer Science, Volume 3911/2006, 970-977.

D.Takahashi (2010), *An Implementation of Parallel 3-D FFT with 2-D Decomposition on a Massively Parallel Cluster of Multi-core Processors*, Parallel Processing and Applied Mathematics, Lecture Notes in Computer Science, Volume 6067/2010, 606-614

P. Wapperom, A.N. Beris, M.A. Straka (2006), *A new transpose split method for three-dimensional FFTs: performance on an Origin2000 and Alphaserver cluster,* Parallel Computing, v. 32, i. 1, pp. 1-13.

T. Weidauer et al.(2010), *Shallow Moist Convection*, John von Neumann Institute for Computing NIC Symposium 2010 Proceedings, 24 - 25 February 2010 : Juelich, Germany, pp. 373-380.